\relax
\documentclass[10pt,twocolumn]{article}
\usepackage{simpleConference}  
\usepackage{times}  
\usepackage{helvet}  
\usepackage{courier}  
\usepackage{url}  
\usepackage{graphicx}  
\usepackage{color}
\usepackage{bbm}
\usepackage{verbatim}
\usepackage{multirow}

\usepackage[numbers]{natbib}

\usepackage{amsmath, amssymb, amsfonts, amsthm}
\usepackage{algorithm, algorithmic}

\theoremstyle{definition}

\usepackage{lipsum}   

\theoremstyle{assumption}

\begin{document}

\title{
Towards Large-Scale Exploratory Search over Heterogeneous Sources}
\author{Mariia Seleznova$^1$, Anton Belyy$^2$, Aleksei Sholokhov$^3$ \\
$^1$ Univ. Grenoble Alpes $^2$ ITMO University, $^3$ University of Washington \\
Email: mselezniova@gmail.com, anton.belyy@gmail.com, aksh@uw.edu
}

\pdfinfo{
/Title (Towards Large-Scale Exploratory Search over Heterogeneous Sources)
/Author (Mariia Seleznova, Anton Belyy, Aleksei Sholokhov)
}

\maketitle

\begin{abstract}
Since time immemorial, people have been looking for ways to organize scientific knowledge into some systems to facilitate search and discovery of new ideas. The problem was partially solved in the pre-Internet era using library classifications, but nowadays it is nearly impossible to classify all scientific and popular scientific knowledge manually. There is a clear gap between the diversity and the amount of data available on the Internet and the algorithms for automatic structuring of such data. In our preliminary study, we approach the problem of knowledge discovery on web-scale data with diverse text sources and propose an algorithm to aggregate multiple collections into a single hierarchical topic model. We implement a web service named \textsc{Rysearch} to demonstrate the concept of topical exploratory search and make it available online.
\end{abstract}

\section{Introduction}
Structuring knowledge and finding relevant literature have always been important problems in science and education. Library cataloging systems \cite{chan2015cataloging} have remained effective search and structuring tools from as early as the fifth century B.C., but recently demonstrated a lack of flexibility dealing with large-scale and diverse documents of the Internet. At the same time, modern search engines, although able to process huge corpora, do not usually provide an overview of knowledge domains and only process short and specific queries. To utilize both approaches, various exploratory search systems \cite{white2009exploratory} have been proposed. They fit well in scientific and popular scientific scenarios, when a user wants to grasp a wide area having no particular request or given an abstract description of several concepts related to some area.   

The integral part of building an exploratory search engine is constructing a representation for documents in the corpus, that can be used for \textit{searching} and \textit{visualization}. Two generic representations are used in NLP, namely 1) long and sparse counter / TF-IDF vectors, as well as 2) short and dense embeddings (such as paragraph2vec or topic models). The latter are more preferable  nowadays due to their compact size and better capturing of synonymy and relatedness.

Among recent approaches for building dense document representations, probabilistic topic models have been shown to perform on par with Skip-Gram Negative Sampling while enjoying improved training performance and interpretability of vector components \cite{potapenko2017interpretable}, which is crucial in large-scale exploratory search scenarios.



Heterogeneous documents on the Internet pose additional challenges for the exploratory search engine design. First, vocabularies used in different Internet sources are often dissimilar and noisy, which might prevent their successful incorporation into a single model. Additional problems occur with \textit{topical imbalance}, when some knowledge domains might become underrepresented in a topic model due to smaller number of documents in the source. Second, the number of concepts discussed in the
documents grows proportionally with the number of documents, and in large-scale scenarios,  the number of topics may become so huge that a user gets confused if all the topics are displayed simultaneously. Turning to hierarchical extentions of topic models (hierarchical topic models or HTMs) might be a beneficial solution in this case.



Our research aims at providing initial answers to these problems and our contribution is therefore twofold: first, we provide an algorithm, which is able to incorporate data into a hierarchical topic model without suppressing underrepresented knowledge domains over web-scale numbers of heterogeneous documents; second, we implement a web service called \textsc{Rysearch} that allows to observe topics and subtopics in a hierarchical \textit{knowledge map}, as well as to perform inexact search over the search index.

We demonstrate the utility of the algorithm and the service on two popular scientific datasets and make our implementations available online.

\section{Aggregation of Heterogeneous Sources}

\begin{table*}
\begin{center}

  \begin{tabular}{ l|c|c|c|c|c|c}
    \hline
    & \multicolumn{2}{|c|}{Average topic quality} & \multicolumn{4}{|c}{Hierarchy edges quality}\\ \hline
     & level 1& level 2 & AP@100 & AP@200 & AP@500 & AP@1000\\ \hline
    No init, no fixed vocab (\textit{Baseline}) & 21.8 & 20.0 & 10.8 & 21.9 & 38.4 & 50.9\\ \hline
    No init, fixed vocab & 23.3 & 19.3 & 37.8 & 42.6 & 48.6 & 58.0\\ \hline
    Init, no fixed vocab & 24.1 & 21.0 & 36.0 & 40.3 & 45.0 & 54.8\\ \hline
    Iterative init, no fixed vocab & 25.8 & 21.8 & \textbf{45.3} & 44.1 & 45.2 & 54.6\\ \hline
    Iterative init, fixed vocab & 27.4 & 21.9 & 44.6 & 45.7 & 49.3 & 57.2\\ \hline
    Init, fixed vocab (\textit{Proposed}) & \textbf{28.4} & \textbf{23.6} & 42.2 & \textbf{47.4} & \textbf{51.7} & \textbf{59.4}\\ 

    \hline
  \end{tabular}
  \caption{\small\label{table:tbl1} Comparison of the baseline algorithm and the proposed modifications. All the proposed modifications show to increase HTM quality on their own and their combination gives the best result. To compare hierarchical topic models, average topic quality based on word embeddings from \cite{Nikolenko2016} for each hierarchy level, as well as ranking quality metrics for topic hierarchy edges from \cite{We2018} are used.}
 \end{center}
\end{table*}

\paragraph{Data} In our experiments we use two datasets gathered from popular scientific websites: Postnauka.ru (around 3000 documents) as the \textit{initial collection} and Habr.com (around 10000 documents) as the collection to add, or \textit{added collection}. Postnauka.ru covers many  major scientific topics, and articles have manually assigned tags set by the website editors, which allow building a model of high interpretability. Habr.com is a blog platform focused on IT.


\paragraph{Baseline Algorithm}
The baseline approach to solve the problem is to merge the collections together and build a new model. As one may see in Table \ref{table:tbl1}, it leads to losing the interpretability at each level of hierarchy, and many smaller topics get lost. The baseline approach can only discover knowledge domains of roughly equal sizes that does not correspond well to the data used, and most of the constructed topics appear to be detailed topics of the bigger source while the topics of the smaller source merge together despite being more interpretable and diverse.

\paragraph{Proposed Algorithm} We propose two modifications to the baseline approach: \textit{initialization} and \textit{fixed vocabulary}.

Initialization is setting the initial parameters of a HTM 
with some approximation before training the model of the merged collection. We use the parameters of an interpretable HTM of the initial collection. As topic modeling problem has multiple solutions, such an approach allows to find a solution that is close to the initial model. The interpretable topics of the initial collection that are not present in the new collection remain almost intact, while the topics specific for the added collection expand and may change their vocabulary accordingly. 

Fixed vocabulary modification prevents the HTM from extending its vocabulary with new words occurring in the added documents. It is equivalent to adding the new words to the stopwords list which is known to improve text analysis quality in the case when the new words are rare. 

We also explore the iterative version of the proposed algorithm: the added collection is separated into batches of gradually increasing size (not larger than 10\% of the merged collection size in the previous iteration) and the next iteration is initialized with the previous iteration model. The results are shown in Table \ref{table:tbl1}.

\paragraph{Implementation}

\begin{figure}
\includegraphics[width=0.5\textwidth]{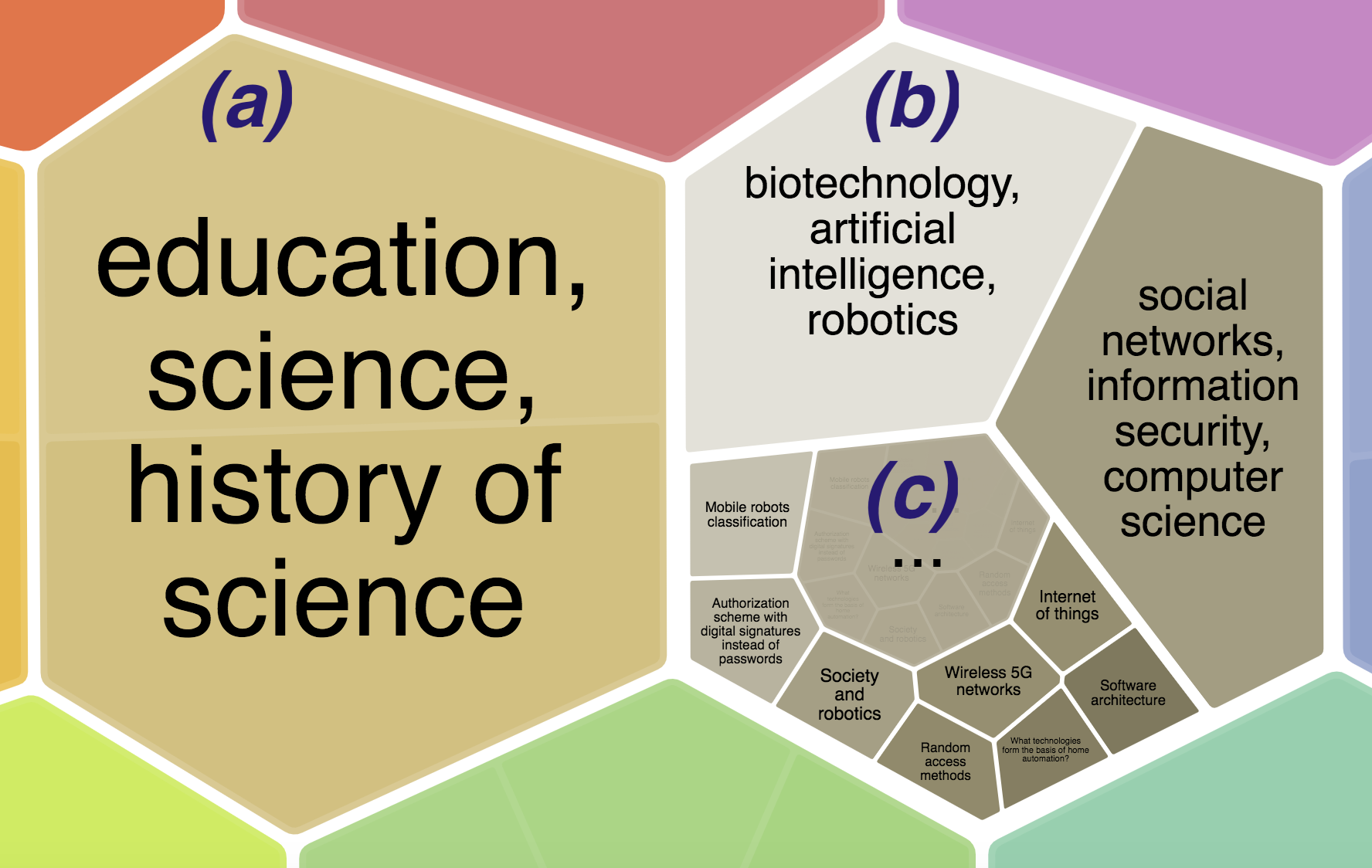}
    \caption{\small\label{fig:lod}Different levels of detail of hierarchical cells on the map: (a) a topic, (b) a topic with three sub-topics, (c) a set of document pertaining to the specific sub-topic, which can be further zoomed in by clicking on a (...) cell.}
\end{figure}

\textsc{Rysearch}\footnote{\url{http://github.com/AVBelyy/Rysearch}} is a web application which provides tools for searching over and exploring heterogeneous hierarchical topic models, represented as an interactive topical map (built with FoamTree visualization library\footnote{\url{https://carrotsearch.com/foamtree/}}). A fragment of such map can be seen on Figure \ref{fig:lod}.


\clearpage
\section{Supplementary material}

\paragraph{Tools} We utilize BigARTM library for efficient construction of topic models. It was shown in \cite{vorontsov2015bigartm} that ARTM implemented in BigARTM outperforms LDA models implemented in state-of-the-art Vowpal Wabbit\footnote{\url{
https://github.com/JohnLangford/vowpal_wabbit/}} and Gensim \cite{rehurek2010software} libraries in terms of training and inference time, as well as on held-out perplexity. Additional benefits of ARTM approach include controlled sparsity of parameters (which helps reducing the size of a search index in exploratory search engines), interpretability of components and easy composability of linguistic priors, or regularizers. For a survey of regularizers, we refer to \cite{vorontsov2014tutorial}. The hierarchical extension of ARTM (or hARTM) that we use is proposed in \cite{Chirkova2016}.

\paragraph{Topic Model Parameters} In our experiments, we build a two-level hARTM model with 21 topics (twenty subject topics, each for a specific knowledge domain, and one for the \textit{background} lexicon, occurring in all the domains simultaneously) at the upper level and 61 topics (sixty subject topics and one background topic) at the lower level, with $20 \times 60 = 1200$ topical \textit{edges} between the pairs of subject topics. We use smoothing and decorrelating ARTM regularizers at both levels to create background topics with common vocabulary and subject topics with diverse and dissimilar vocabulary.

\paragraph{Quality Evaluation}
In the paper, we propose two modifications to the baseline algorithm, namely \textit{initialization} and \textit{fixed vocabulary}. We then perform ablation study, where we show that each modification individually improves model's quality, and their combination works the best. Here we report more results related to the same ablation study. First, let us define the following short names for the proposed \textit{combinations} of modifications:

\begin{itemize}
\item \textbf{D- I-} is the baseline algorithm, with non-fixed {d}ictionary and no {i}nitialization.

\item \textbf{D+ I-} has fixed {d}ictionary, but no {i}nitialization.

\item \textbf{D- I+} has non-fixed {d}ictionary with {i}nitialization.

\item \textbf{D+ I+} is the proposed algorithm, with both fixed {d}ictionary and {i}nitialization.

\item \textbf{D+ I+-} is the \textit{iterative} version of the proposed algorithm, where the added collection is merged into the topic model in a succession of small batches.

\item \textbf{D- I+-} is the \textit{iterative} version of the proposed algorithm with non-fixed dictionary.
\end{itemize}

\begin{figure}
\includegraphics[width=0.5\textwidth]{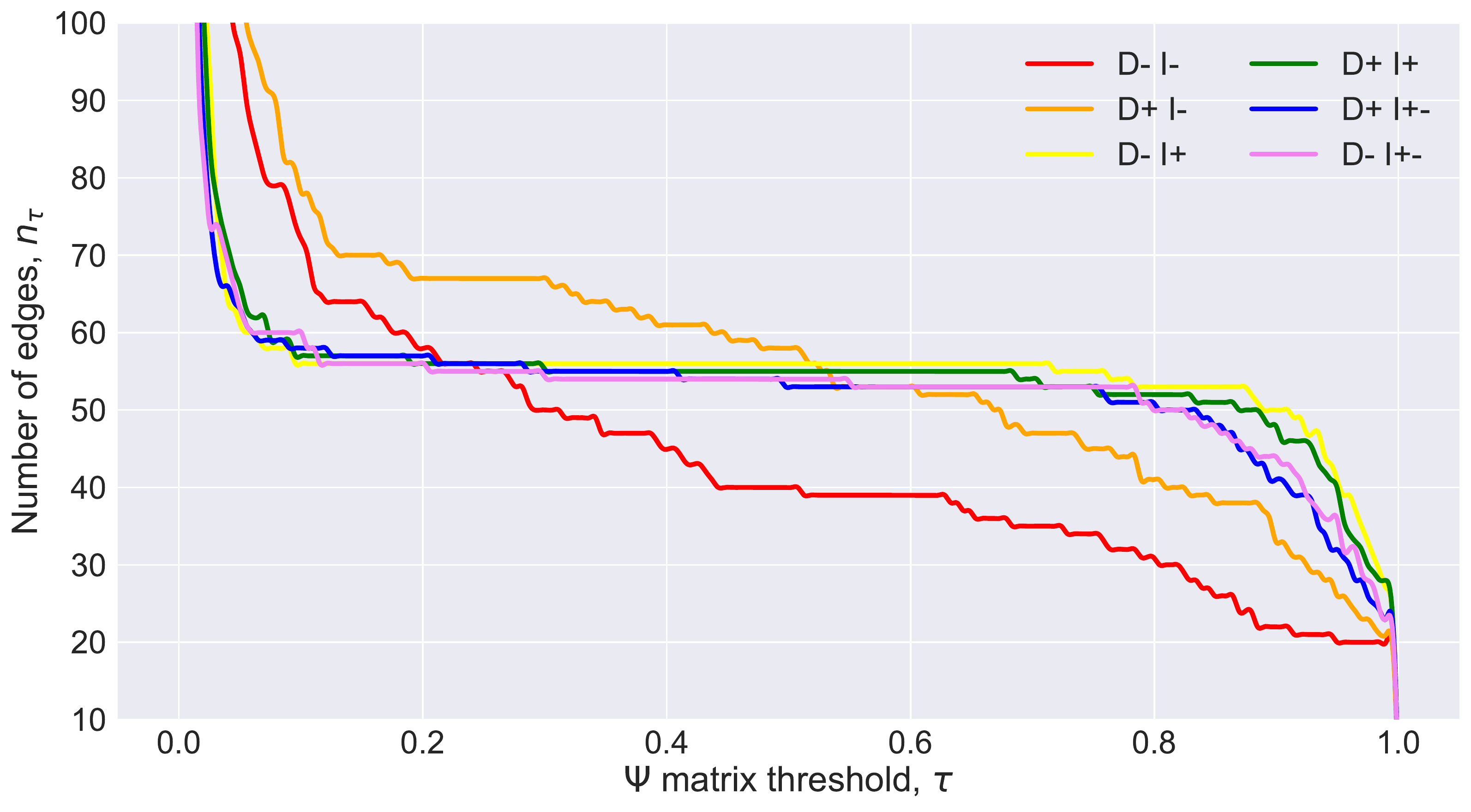}
    \caption{\small\label{fig:num_edges} Number of edges $n_\tau$ in each combination, calculated as $n_\tau = \sum_{a, t} [\Psi_{at} \geq \tau]$ where $\Psi$ is the model's parameter matrix.}
\end{figure}

The number of edges depends on the application and can be chosen using Figure \ref{fig:num_edges}. We can observe that for better performing combinations the line is ``steeper'' in the areas of low and high probabilities and reaches a plateau in the middle, meaning that edges of high and low probability are well-separated, so the number of edges plot might also serve as a simple proxy to the quality of a topical hierarchy. 

\begin{figure*}
\includegraphics[width=0.95\textwidth]{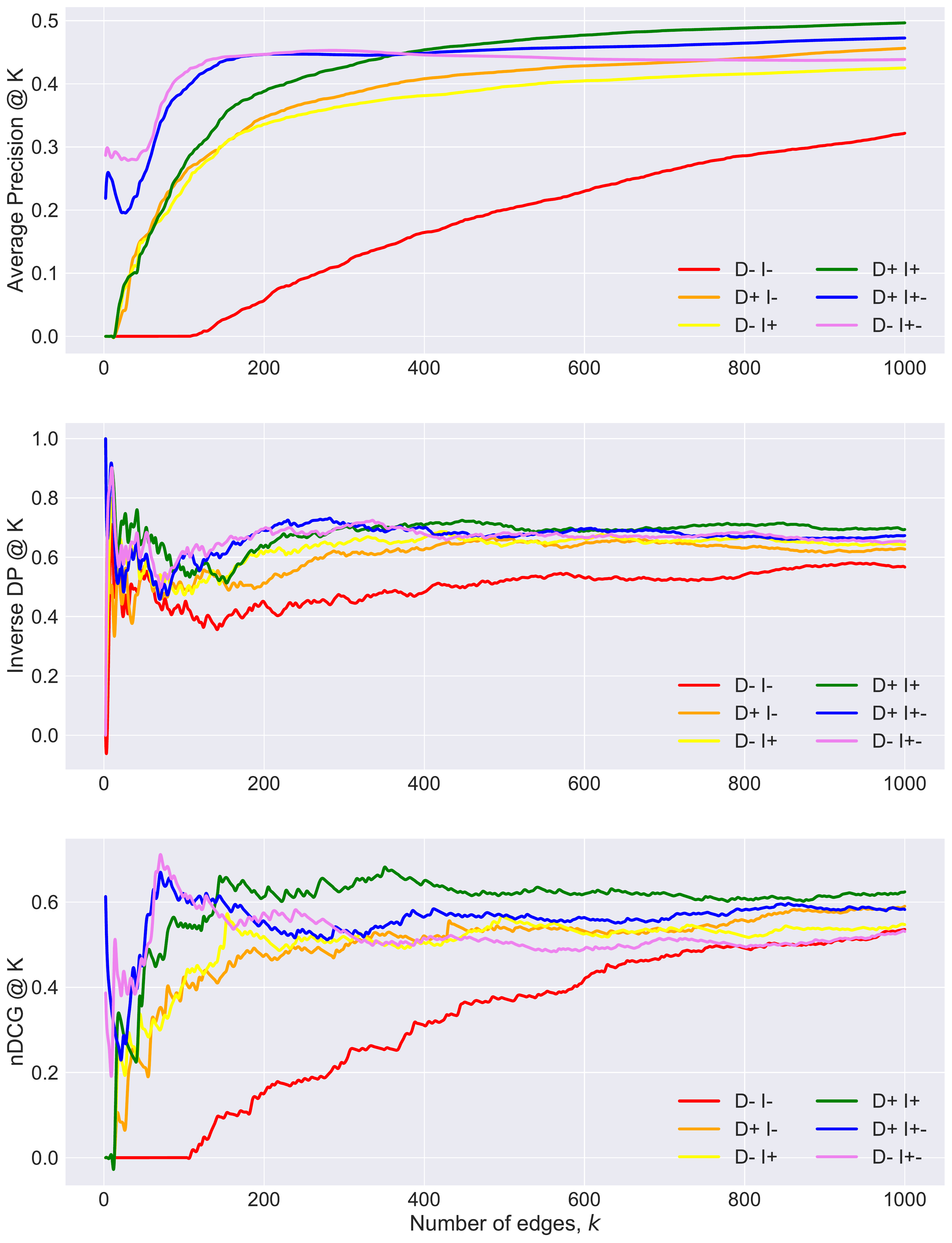}
    \caption{\small\label{fig:ranking_stats} Ranking quality measures for each combination's topical hierarchy.}
\end{figure*}

In Figure \ref{fig:ranking_stats} we provide ranking quality measures calculated with embedding similarity between each pair of topics' top 10 words to show how well the model ranks the hierarchy edges from the best to the worst ones. These metrics depend on the number of edges and are explained in more detail in \cite{We2018}.

\clearpage

\bibliography{references}
\bibliographystyle{plainnat}

\end{document}